\documentclass[aip,jcp,reprint,superscriptaddress,numerical]{revtex4-2}

\usepackage{amsmath}
\usepackage{amssymb}
\usepackage{graphicx}
\usepackage{bm}
\usepackage{dcolumn}
\usepackage{hyperref}
\usepackage{xcolor}
\hypersetup{colorlinks=true, linkcolor=blue, citecolor=blue, urlcolor=blue}
\begin{document}

\title{Correlated-Electron Theory of Triplet-Triplet Multiexciton States in Polypentacene}

\author{Rupali Jindal}
\affiliation{Department of Physics, University of Arizona, Tucson, Arizona 85721, USA}

\author{Alok Shukla}
\affiliation{Department of Physics, Indian Institute of Technology Bombay, Powai, Mumbai 400076, India}

\author{Sumit Mazumdar}
\email{mazumdar@arizona.edu}
\affiliation{Department of Physics, University of Arizona, Tucson, Arizona 85721, USA}

\date{June 15, 2026}

\begin{abstract}
	We present correlated-electron calculations of optical spin-singlet and triplet-triplet multiexciton states in three- and four-unit pentacene oligomers as microscopic models for polypentacene.  The calculations use the Pariser-Parr-Pople Hamiltonian, multiple-reference singles and doubles configuration interaction, and a molecular exciton basis that resolves Frenkel, charge-transfer, and triplet-pair (T$_1$T$_1$) configurations in real space.  We find that the complete set of $^1$(T$_1$T$_1$) eigenstates lies in a narrow, nearly degenerate energy window near the lowest optical exciton and that no eigenstate can be identified with a single localized triplet-pair configuration.  Instead, each triplet-pair eigenstate is a quantum superposition of configurations containing all accessible intertriplet separations.  This electronic structure explains the perceived absence of intramolecular triplet diffusion in pentacene oligomers, polypentacene, and polytetracene solutions, while leaving open the possibility of intermolecular singlet fission in films with appreciable interchain interactions.
\end{abstract}

\maketitle

\section{Introduction}

Singlet fission (SF) is the spin-allowed conversion of the lowest optically allowed singlet exciton S$_1$ into a spin-correlated triplet pair, $^1$(T$_1$T$_1$).  Because subsequent separation of this triplet-pair multiexciton can in principle yield two independent T$_1$ excitons from one absorbed photon, SF continues to be examined as a route to increasing the photocurrent available from molecular and polymeric semiconductors \cite{Smith13a,Rao17a}.  The energetic criterion E(S$_1$) $\geq$ 2E(T$_1$) is necessary, but not sufficient, for useful SF: the initially formed correlated triplet pair must also escape geminate recombination and evolve into triplets that are electronically and spatially distinct.  The nature of the $^1$(T$_1$T$_1$) eigenstates, and not merely their energies, is therefore central to understanding whether a singlet-fission material can support triplet separation.

Covalently linked acene oligomers provide a particularly direct setting in which to examine this problem.  Intramolecular singlet fission (iSF) has been reported in a wide range of pentacene and tetracene dimers and higher oligomers, where the chromophore identity, connectivity, and bridge length can be varied systematically \cite{Sanders15a,Zirzlmeier15a,Lukman15a,Sakuma16a,Sanders16a,Korovina16a,Korovina18a,Zirzlmeier16a,Sun16a,Basel17a,Sakai18a,Miyata19a,Musser19a,Sakai26a}.  These systems have also exposed an unresolved issue: the experimentally observed triplet-pair state is often described as a localized, bound nearest-neighbor multiexciton, whereas a conjugated oligomer can support coherent mixing among triplet-pair configurations with different interchromophore separations.  Distinguishing between these pictures is essential, because a localized-pair model invites discussion in terms of a distance-dependent binding energy E$_b$, while a delocalized or configuration-mixed eigenstate may not admit such a simple interpretation.

The same question has a close connection to the older correlated-electron description of $\pi$-conjugated polymers.  In linear polyenes, the optically dark 2$^1$A$_g$ state lies below the dipole-allowed 1$^1$B$_u$ state for sufficiently long chains \cite{Hudson74a,Hudson82,Kohler88a}.  Calculations based on the Pariser-Parr-Pople (PPP) 
Hamiltonian \cite{Pariser53a,Pople53a} showed that this state has predominantly covalent two-triplet character, in contrast to the more ionic optical exciton \cite{Schulten76a,Tavan79a,Ramasesha84b,Ramasesha84c,Tavan87a}.  Higher two-triplet states have also been identified in correlated treatments of polyenes \cite{Valentine20a,Wang21b,Manawadu22a}.  We use this historical connection only as a conceptual guide: the present acene oligomers differ from ideal polyenes in monomer size, excitation localization, and the role of intermonomer charge transfer, but both classes of systems demonstrate that the two-triplet sector is intrinsically a many-electron, strongly correlated problem.

Recent experiments on polypentacene (PPc), finite pentacene oligomers $n$Pc, and polytetracene (PTc) were motivated by the possibility that longer covalently connected acene chains might reduce the effective triplet-pair binding and permit rapid intramolecular triplet separation \cite{Sanders16d,Pun17a}.  In pentacene dimers, the $^1$(T$_1$T$_1$) state is strongly bound and triplet separation is inefficient \cite{Sanders15a}.  It was natural to expect that increasing the number of acene units would open additional spatial configurations and lower the energetic cost of triplet separation.  Transient absorption (TA) measurements, however, continued to observe absorption from nearest-neighbor triplet pairs in longer oligomers and in PPc, leading to the interpretation that the triplets remain confined by a large binding energy \cite{Sanders16d}.  This interpretation implicitly assumes that individual triplet-pair configurations, such as those sketched in Fig.~1(a), are approximate eigenstates and that triplet separation proceeds through discrete jumps between them.

Here we show that this localized-configuration picture requires substantial revision.  Full correlated-electron calculations for 3Pc and 4Pc demonstrate that the $^1$(T$_1$T$_1$) eigenstates are not single nearest-neighbor, second-neighbor, or distant triplet-pair configurations.  Rather, the triplet-pair manifold consists of a nearly degenerate set of quantum eigenstates, each of which contains coherent contributions from configurations with multiple intertriplet separations, as represented schematically in Fig.~1(b).  Consequently, the persistence of TA features associated with nearest-neighbor triplet pairs does not by itself establish a large triplet-triplet binding energy in long acene oligomers or PPc.  The more appropriate conclusion is that intrachain triplet separation into a unique, energetically isolated distant-triplet eigenstate is not supported by the correlated electronic structure.

Our calculations are performed with the PPP Hamiltonian and multiple-reference singles and doubles configuration interaction (MRSDCI).  To obtain physical wavefunction assignments, we use a diagrammatic molecular exciton basis that separates intramonomer Frenkel excitations, intermonomer charge-transfer configurations, and $^1$(T$_1$T$_1$) multiexciton configurations with distinct intertriplet distances \cite{Khan17b,Khan18a,Parenti23a,Chesler24a,Nazir25a}.  This basis makes it possible to compare directly the localized-pair and configuration-mixed descriptions, to compute ground-state absorptions and excited-state absorptions from triplet-pair eigenstates, and to reinterpret the transient absorption experiments on $n$Pc, PPc and PTc within a unified correlated-electron framework.

\begin{figure}[h]
\centerline{\resizebox{3.0in}{!}{\includegraphics{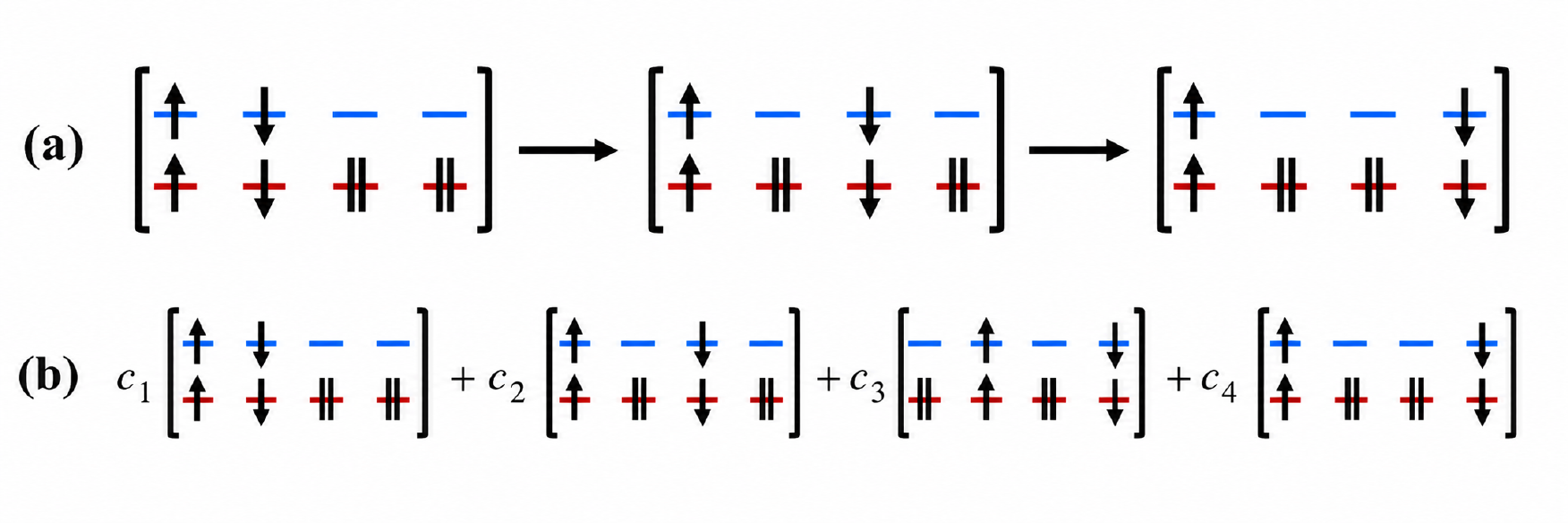}}}
\caption{Schematics of two different conceptualizations of triplet-triplet
eigenstates of 4Pc (see text). Red and blue horizontal bars correspond to HOMO and LUMO of
Pc monomers, doubly or singly occupied by electrons. Arrows on singly occupied MOs indicate electron spins. Bonding (antibonding) MOs
not shown are completely filled (empty). (a) Individual configurations are eigenstates,
	and triplet separation occurs along the indicated path. (b) Eigenstates are quantum superpositions of configurations with different relative weights $c_i$.}
\end{figure}

\section{Results and Discussion}

\subsection{Pentacene Oligomers}

We have investigated theoretically the excited state electronic structures
of 3Pc and 4Pc, with \textit{anti-anti} and \textit{anti-anti-anti}
intermonomer bonding motifs (see Fig.~2). The $\text{C}\equiv\text{C}$ groups were included in our calculations in order to take into account the TIPS-pentacene polymers investigated experimentally\cite{Sanders16d}. Our earlier theoretical works have shown
that in contrast to Pc oligomers with short bridge molecules (anthracene
or shorter) \cite{Parenti23a}, the optical behavior of oligomers with longer bridge molecules is independent of bonding motif \cite{Chesler24a,Nazir25a}. The latter was found to be true also experimentally for $n$Pc \cite{Sanders16d}.
We have therefore not investigated structures with \textit{syn}-bonding.
Wherever necessary, we compare our computational results with those
obtained earlier for 2Pc \cite{Khan17b}.
\begin{figure}[h]
\centering{}\includegraphics[width=\columnwidth]{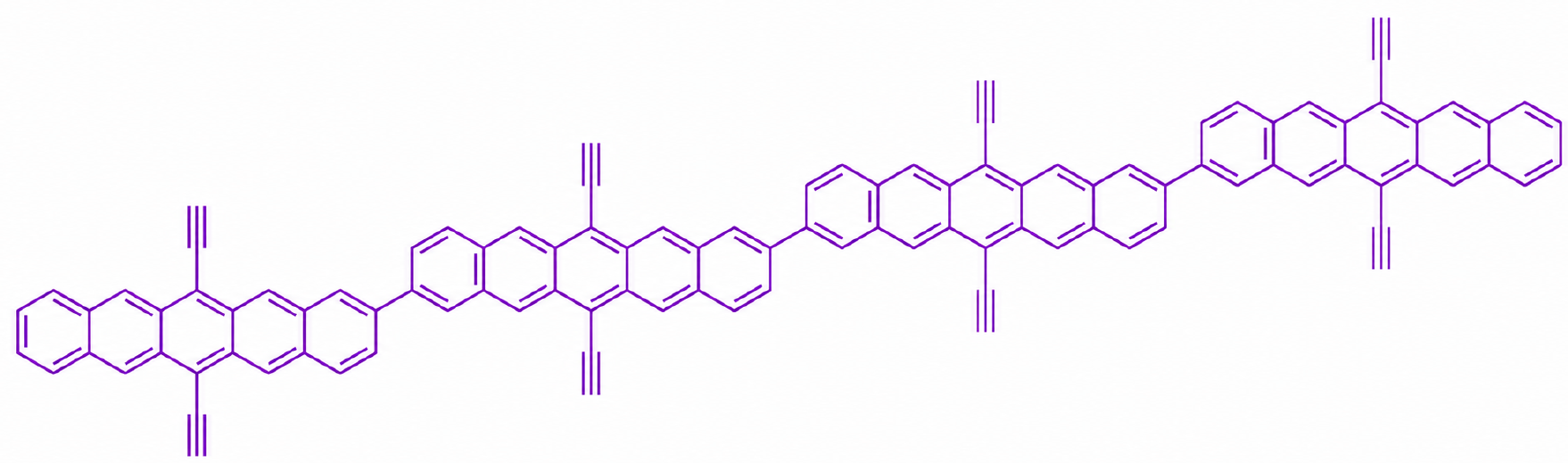}
\caption{Pentacene tetramer 4Pc with the anti-anti-anti intermonomer bonding motif.}
\end{figure}

\subsection{PPP Hamiltonian, MRSDCI, and Diagrammatic Exciton Basis}

As has been demonstrated with polyenes \cite{Schulten76a,Tavan79a,Ramasesha84b,Ramasesha84c,Baeriswyl92a,Soos94a,Ramasesha00a,Guo94a,Barford05a,Mazumdar09a}, and
in our previous work on acene iSF oligomers \cite{Khan17b,Khan18a,Parenti23a,Chesler24a,Nazir25a}, precise theoretical
understanding of the two electron-two hole (2e-2h) $^{1}$(T$_{1}$T$_{1}$)
excitation necessarily requires incorporation of configuration interaction
(CI) with up to quadruple (4e-4h) excitations from the mean-field Hartree-Fock ground state, as well as the inclusion of the largest number of
molecular orbitals (MOs) per monomer. Incorporating such high-order CI remains   
out of bounds within first-principles based approaches for systems with more than 20 $\pi$-electrons \cite{Epifanovsky21a} (which is close to one-quarter of the number of $\pi$-electrons in 4Pc), for the simple reason that the number of 4e-4h excitations explodes with system size. 
This problem is further 
compounded in the present case, because we will be interested in (i) determining the energies and wavefunctions of the complete set
of covalent $^1$(T$_1$T$_1$) states, as opposed to only the lowest one, and (ii) calculating excited state absorptions (ESAs) {\it from} the $^1$(T$_1$T$_1$)
states in order to understand the experimental TA data. The final states of the TAs are at even higher energy than the $^1$(T$_1$T$_1$) and their
computations not only require great precision, but once again, inclusion of large number of MOs per monomer.
AS in our previous
work on shorter acene oligomers \cite{Khan17b,Khan18a,Parenti23a,Chesler24a,Nazir25a}, our calculations are therefore
within the semiempirical PPP Hamiltonian, written as,

\begin{multline}
H=\sum_{\langle ij\rangle,\sigma}t_{ij}(c_{i\sigma}^{\dagger}c_{j\sigma}+c_{j\sigma}^{\dagger}c_{i\sigma})+U\sum_{i}n_{i\uparrow}n_{i\downarrow}\\
+\frac{1}{2}\sum_{i\neq j}V_{ij}(n_{i}-1)(n_{j}-1)
\end{multline}

\noindent Here, $c_{i\sigma}^{\dagger}$ creates an electron with spin
$\sigma$ on the $p_{z}$ orbital of a C atom, $n_{i\sigma}=c_{i\sigma}^{\dagger}c_{i\sigma}$
is the number of electrons with spin $\sigma$ on atom $i$, and $n_{i}=\sum_{\sigma}n_{i\sigma}$
is the total number of electrons on the atom. The symbol $\langle\rangle$
refers to nearest-neighbor C atoms and $t_{ij}$ are the corresponding
electron hopping integrals, $U$ the Coulomb repulsion between two
$\pi$ electrons occupying the same $p_{z}$ orbital, and $V_{ij}$
the long-range Coulomb interaction. The parameters of the Hamiltonian were chosen from extensive comparisons to experiments on acene
monomers and dimers, as discussed in our earlier works \cite{Khan17b,Khan18a}. Peripheral and internal acene C-C bond lengths (hopping integrals)
are taken to be 1.40 \AA (-2.4 eV) and 1.46 \AA (-2.2 eV), respectively. For the TIPS group, the C--C single and triple bond lengths (hopping integrals) are taken to be 
1.46 \AA{} (-2.2 eV) and 1.20 \AA{} (-3.0 eV), respectively\cite{ducasse1982correlated}.
We have chosen $V_{ij} = U/\kappa(1 + 0.6117R_{ij}^2)^{1/2}$, where $R_{ij}$ is the distance in \AA~ between C-atoms $i$ and $j$ and $\kappa$ an 
effective dielectric constant. The onsite Hubbard repulsion $U$ and the dielectric constant $\kappa$ are taken to be 6.7 eV and 1.0 based 
on fitting monomer singlet and triplet energies \cite{Khan17b,Khan18a}. 
Electron-vibration interactions are not
included; simultaneous inclusions of the CI with quadruple excitations that
is essential for even qualitative understanding of the $^{1}$(T$_1$T$_1$)
states and electron-vibration interactions are outside the scope of
current many-body computational approaches.

In order to incorporate CI with the most relevant 4e-4h quadruple excitations
we have used the multiple reference singles and doubles CI approach
\cite{Tavan87a}, which has been discussed extensively in our recent
work \cite{Khan17b,Chesler24a,Nazir25a,Khan18a}. 
The approach essentially involves including CI with 2e-2h double excitations from the most relevant 2e-2h double excitations that describe
a targeted excited state, in a stepwise manner until convergence is reached  (see SM, Section III). The {\it same} procedure is followed in our calculations of ESAs from the 
$^1$(T$_1$T$_1$) eigenstates, where the final eigenstates are dipole-coupled to the initial $^1$(T$_1$T$_1$). 

Beyond MRSDCI, our computational approach is unique in the use of a many-electron molecular exciton basis space. The standard  MO-based approach to CI is not suitable for our purpose, as our goal is
to obtain physical, pictorial descriptions of excited states (in particular
the $^{1}$(T$_1$T$_1$) state) that allow locating the excitations in configuration
space (in order to distinguish between the configurations of Fig.~1), and identifying their natures (\textit{i.e.}, 
1e-1h versus 2e-2h; intra- versus intermonomer,
etc.). Our calculations therefore are within the diagrammatic molecular
exciton basis \cite{Khan17b,Khan18a,Chesler24a,Nazir25a}. In the present work,
we have retained 10 frontier MOs per pentacene monomer (5 bonding
and 5 antibonding), \textit{i.e.}, 30 and 40 MOs for 3Pc and 4Pc,
respectively. These are the largest CI calculations of 
$^1$(T$_1$T$_1$) states to date. The dimensions of the MRSDCI matrices
in all cases run into multiple millions (see SM III).

\subsection{Oligomer Geometry}

The motivation of our semi-empirical calculations is to arrive at correct physical interpretations of the nature of the triplet-triplet eigenstates and TA spectroscopy that
have been used to identify the same, and not quantitative fitting. We have done our calculations for two different oligomer geometries, viz., (i) planar, and (ii) with dihedral
angles 30$^o$ between the Pc monomers. In the main text we report computational results for 3Pc and 4Pc with planar geometries only. Computationl results 
corresponding to (ii) are presented in detail in SM VI. The $^1$(T$_1$T$_1$) wavefunctions and ESA spectra in the two cases are very similar. The natures of the 
relevant wavefunctions therefore depend weakly on the torsion angle \cite{Khan17b}. The iSF rate, however can be influenced 
by torsional angle.

\subsection{Ground-State Absorption}

Ground-state optical absorption spectra for $n$Pc,
$n=2-4$ were calculated using the standard Lorentzian expression for absorption, PPP-MRSDCI excited state energies and dipole couplings, 
and uniform linewidth parameters 0.1 eV. The calculated absorption spectra are shown in Fig.~3. Table I gives the energies, transition dipole couplings with the correlated ground state, and the
oscillator strengths of the final eigenstates that constitute the absorption bands labeled S$_1$, S$_2$, S$_3$ and S$_4$ in Fig.~3. 
The calculated absorption spectra reproduce the experimental electronic absorption bands very well, except for weak relatively
blue shifted energies of the calculated bands, see Fig.~4 of Supplementary Information (SI) of  
Ref.~\onlinecite{Sanders16d}. 

Absorption bands S$_1$, S$_2$, S$_3$ and S$_4$ in Fig.~3 are dominated by distinct classes of many-electron configurations. Dominant exciton basis contributions to these final states are shown in Figs.~4(a) and (b), respectively, for $n$ = 3 and 4. We have not shown the corresponding wavefunctions explicitly for $n=2$,
as the S$_1$ and S$_2$ absorption bands there and the corresponding final
eigenstates were discussed extensively in a previous work \cite{Khan17b}. The highest energy absorptions are predominantly intramonomer and are of marginal interest in the context of iSF. Observations on the computational results follow.

\begin{figure}[h]
\centering{}\includegraphics[scale=0.33]{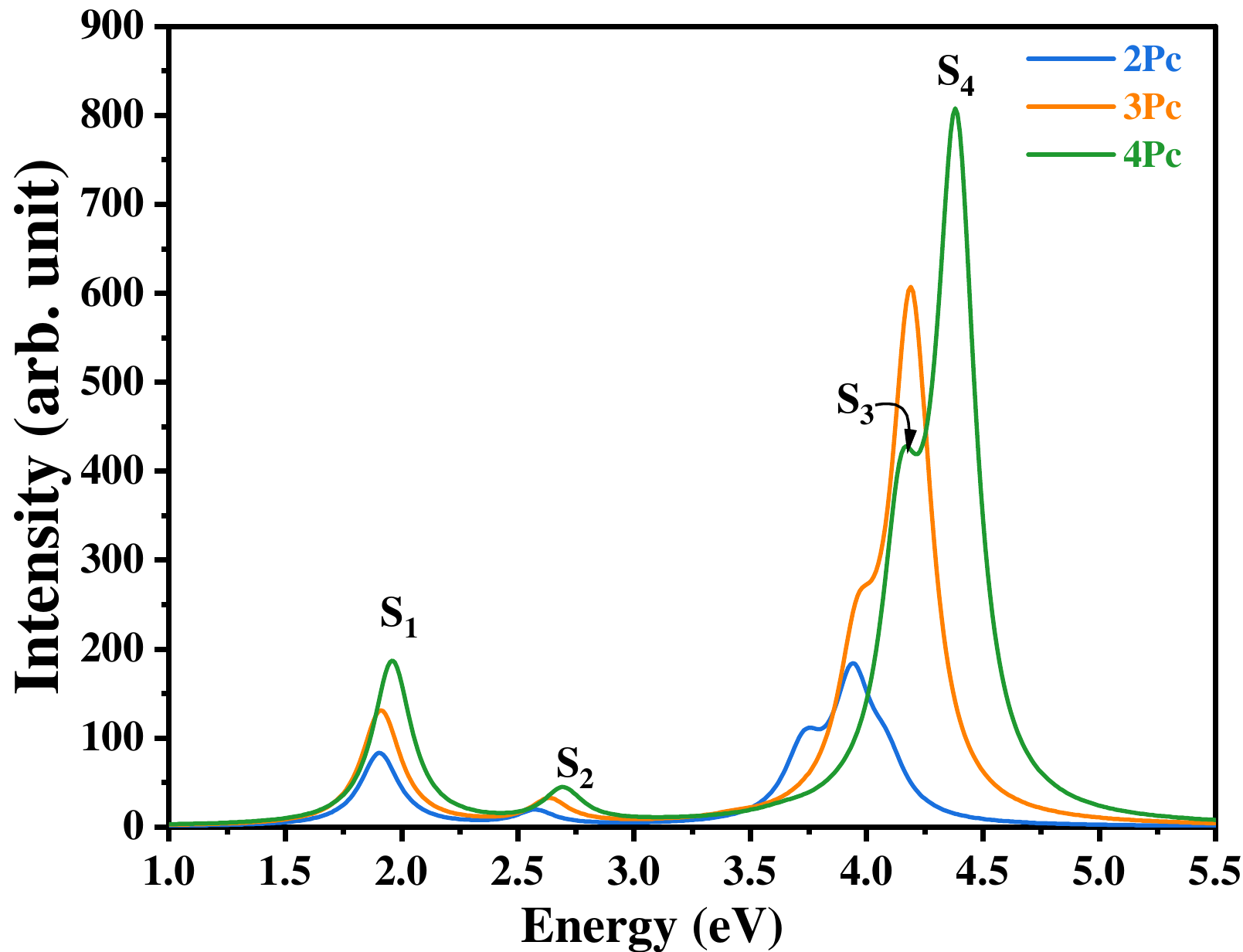}
\caption{ Calculated ground state absorption spectra
	of $n$Pc, $n=$ 2-4. S$_i$, $i$=1-4, label the different absorption bands (see Fig.~4)}
\end{figure}

1) Experimentally, the lowest energy absorption band S$_1$ is accompanied by vibrational sidebands.
The sidebands are not expected within the purely electronic PPP Hamiltonian of Eq.~(1). 

\begin{figure*}[t]
\centering
\includegraphics[scale=0.58]{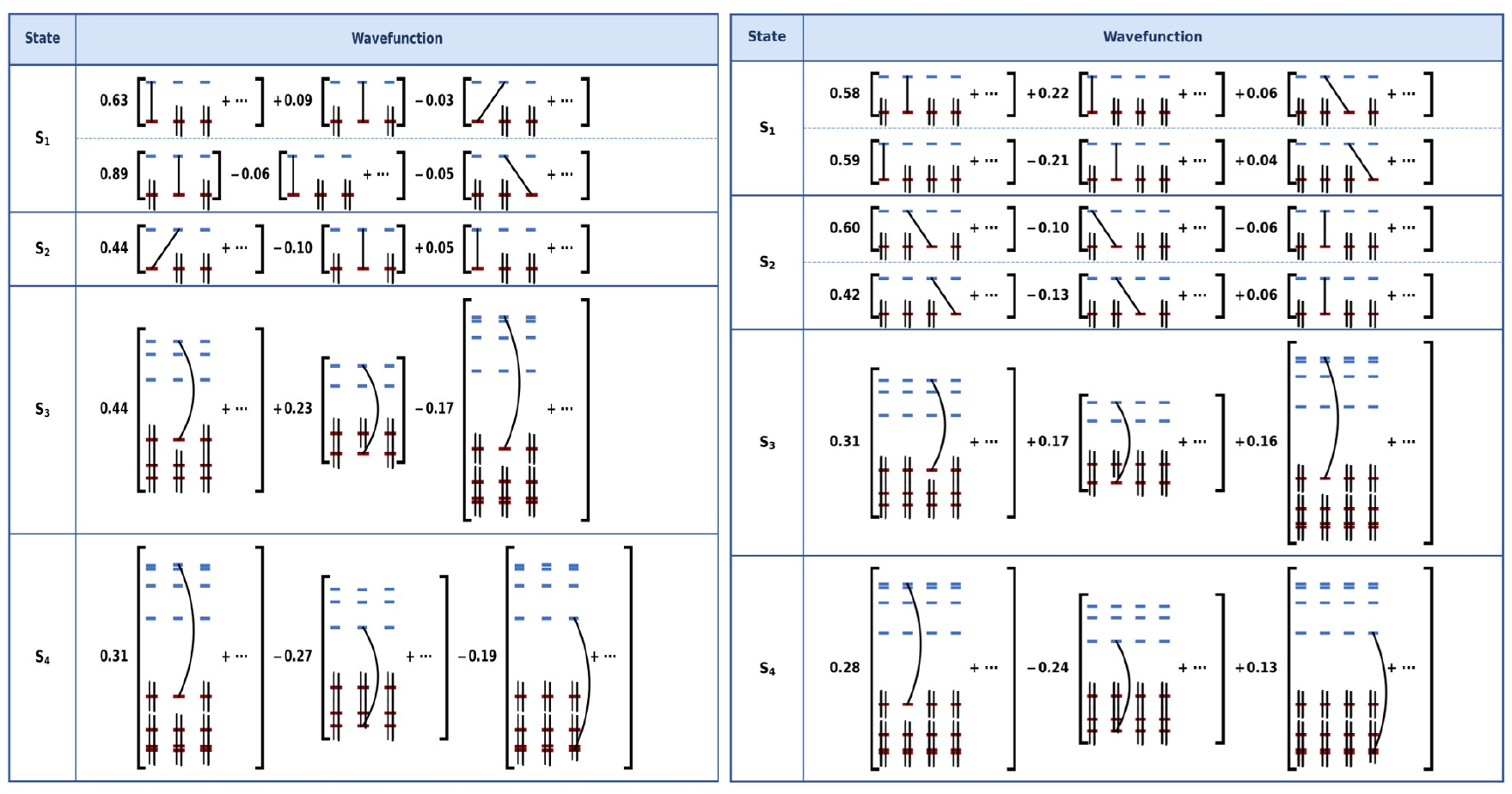}
	\caption{Normalized exciton basis wavefunctions of final eigenstates to which ground state absorption occurs in 3Pc (left) and 4Pc (right).
S$_1$ - S$_4$ correspond to the absorption bands labeled in Fig.~3. Red and blue horizontal bars represent bonding and antibonding MOs, respectively,
which can be doubly or singly occupied by electrons. Straight or curved lines connecting MOs represent spin-singlet excitations. Ellipses correspond to
	configurations related by spatial and charge-conjugation symmetries. Filled (empty) bonding (antibonding) MOs are not shown.}
\end{figure*}

2) Absorption band S$_1$ in $n>$1 is due to transitions to multiple nearly degenerate eigenstates that are predominantly Frenkel exciton states with weak but nonzero intermonomer CT. 

3) We expect and find three and four independent Frenkel exciton basis configurations in 3Pc and 4Pc, respectively. The two dipole-allowed Frenkel eigenstates in 3Pc are predominantly (i) superposition of excitations on the terminal monomers, and (b) excitation on the inner monomer
(with weaker contributions from inner and outer monomers, respectively),
see Fig.~4(a). Similar superpositions are seen in 4Pc, with even
and odd superpositions of excitations on terminal and  central monomers
(see Fig.~4(b)), with the difference between the transition dipole
couplings now significantly larger than in 3Pc (see Table~\ref{tab:energy_dipole_P3_P4}). Dipole-forbidden eigenstates 
are not included in the Figures. 

\begin{table*}[t]
    \centering
    \caption{\label{tab:energy_dipole_P3_P4} Energies, transition dipole moments $\mu_x$ and $\mu_y$ (in \AA, electronic charge $e = 1$) and oscillator strengths $f$ of optical transitions from the ground states in 3Pc (left) and 4Pc (right). Here $x$ and $y$ refer to directional vectors longitudinal and transverse to the monomer axes.}
    \includegraphics[scale=0.55]{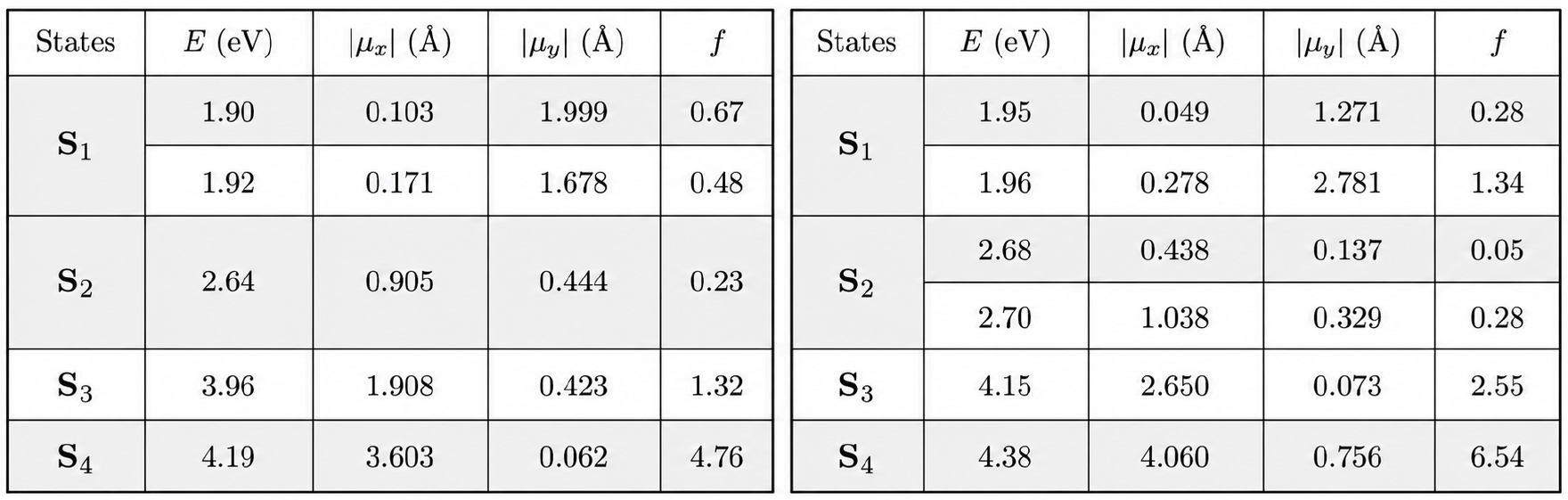}
\end{table*}

4) The next higher energy dipole-allowed absorptions (labeled S$_2$ in Fig.~3) are to CT states.
The weak but observable increase in the calculated oscillator strengths with increasing $n$ is in agreement with the experimental
observation \cite{Sanders16d}.

5) The highest energy intense absorption bands (S$_{3}$ and S$_4$ in Figs.~4(a) and (b)) are due to transitions predominantly from lower single-particle energy monomer bonding MOs to higher single-particle energy monomer antibonding MOs. Unlike the
lowest Frenkel exciton state, the transition dipole vectors corresponding to the high energy absorptions are directed along the longitudinal axis of the oligomers (see Table~\ref{tab:energy_dipole_P3_P4}) and hence their strengths change more rapidly with system size.

\subsection{$^1$(T$_1$T$_1$) Eigenstates}

We now discuss the 2e-2h $^1$(T$_1$T$_1$) eigenstates in $n$Pc, $n=3$ and 4, with emphases on
both similarities and differences with the corresponding states in polyenes 
\cite{Schulten76a,Tavan79a,Ramasesha84b,Ramasesha84c,Tavan87a,Valentine20a,Wang21b,Manawadu22a}. 
Since the triplet excitons in the Pc monomers occur below half the energy of the monomer optical
exciton, and since T$_1$ delocalization is expected to be smaller than the delocalization of
S$_1$, we expect all possible $^1$(T$_1$T$_1$) eigenstates to occur below the lowest S$_1$.
Their total number should be $^n$C$_2$.

Figure~5 shows the three $^1$(T$_1$T$_1$) eigenstates that occur in 3Pc. The calculated energies of the eigenstates are included. Their occurrences at or below the S$_1$ state energies (Table I) demonstrates that the MRSDCI approach, combined with a large number of MOs per monomer, successfully reproduces the experimental results.
Broadly speaking, the wavefunctions can
be thought of as (i) predominantly even and odd linear superpositions of nearest neighbor triplet-triplets, with weaker
contribution by the second neighbor triplet-triplet to the former, and (ii) predominantly second neighbor triplet-triplet
with weak contribution from the even superposition of nearest neighbor triplet-triplets. 

Six distinct $^1$(T$_1$T$_1$) configurations are expected in 4Pc, consisting of three, two and one nearest-neighbor, second neighbor and maximally distant triplet-triplet, respectively. The six eigenstates, which are superpositions of all six configurations in every case, are shown in Fig.~6. Three of the six are dominated by configurations with nearest-neighbor triplet-triplets, with more distant triplet-triplets making weaker contributions. 

\begin{figure}[h]
\centering{}\includegraphics[scale=0.13]{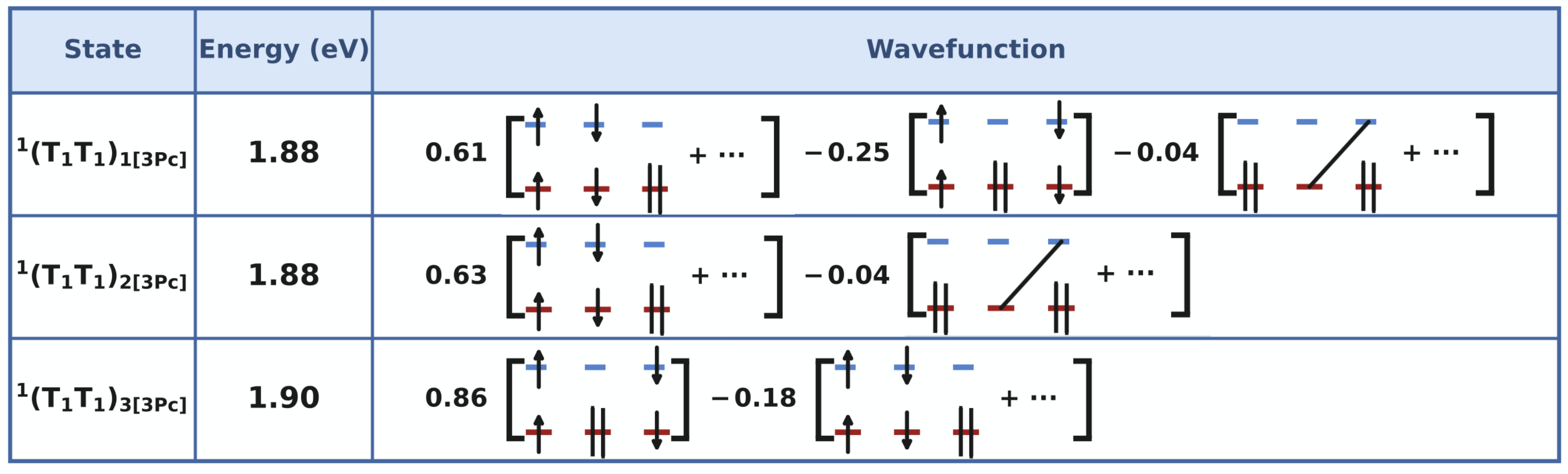}
\caption{\label{fig:TT-P3} $^{1}$(T$_1$T$_1$) eigenstates of 3Pc.}
\end{figure}

The other three consist of configurations with predominantly distant triplet-triplets. The contributions by 1e-1h CT configurations are very small in all cases.

We make the following additional observations.

1) The $^1$(T$_1$T$_1$) eigenstates are very different from what is implicitly assumed in Refs.~\onlinecite{Sanders16d} and \onlinecite{Pun17a} (single configurations as 
eigenstates, see Fig.~1(a)). 
In all cases, the eigenstates are superpositions of configurations with all possible intertriplet distances. Furthermore, the energy differences between the eigenstates
are tiny. The concept of triplet-triplet binding energy does not apply to $n$Pc.

(2) There is an overall similarity with long polyenes in that the number of $^1$(T$_1$T$_1$) eigenstates increases with oligomer length. However, the difference between the two classes of systems is stronger in that there is in $n$Pc a clear energy separation between the entire collection of covalent 
2e-2h triplet-triplets and the 1e-1h CT states S$_2$. The latter is not true in the polyenes, for the reason that the ``monomer'' of 2 C-atoms
there is much too small.

\begin{figure}[h]
\centering{}\includegraphics[scale=0.35]{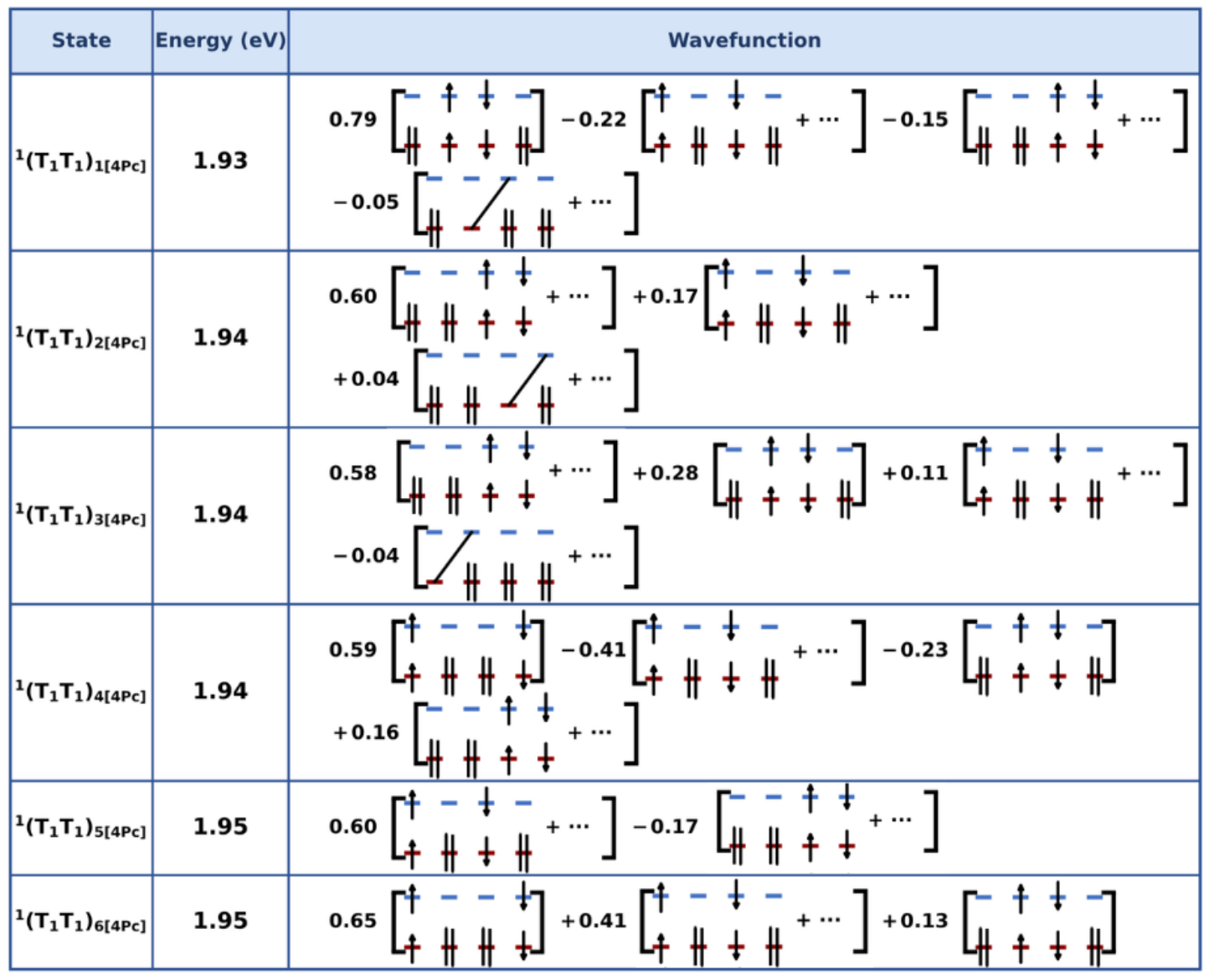}
\caption{\label{fig:TT-P4} $^{1}$(T$_1$T$_1$) eigenstates of 4Pc.}
\end{figure}

(3) The very weak contributions of CT configurations are also in strong contrast to the even parity eigenstates in the polyphenylenes, in which the relative contributions by 1e-1h and 2e-2h configurations are nearly equal and all even parity eigenstates occur above the optical S$_1$ \cite{Chakrabarti99a}.

\subsection{Transient Absorption and ESAs from Triplet-Pair States}

In TA measurement the pump pulse first creates population in the initial excited state, and the delayed probe pulse promotes this already excited state population
to yet higher energy states. The probe absorption energies and durations therefore provide spectroscopic fingerprints of the initial excited state. 
TA spectra of solutions of 2Pc, 3Pc, 4Pc, 5Pc and PPc in 1,2,4-trichlorobenzene (TCB) have been reported in Ref.~\onlinecite{Sanders16d} (see Fig.~3 in SI of Ref.~\onlinecite{Sanders16d}). The samples were excited at 600 nm,
and the TA spectra were obtained 10 ps after excitation. Over and above the high energy TAs at $\sim$ 500-525 nm that correspond to intramonomer T$_1$ excitation (as determined
by comparison to TA spectra of photosensitized triplets), strong photoinduced absorptions at 710 nm that were absent in the photosensitized triplets were observed 
from the $^1$(T$_1$T$_1$) states in all cases. This led to the idea that the $^1$(T$_1$T$_1$) binding energy of $n$Pc and PPc are high, as we describe below.

\begin{figure}[h]
\centering{}\includegraphics[scale=0.20]{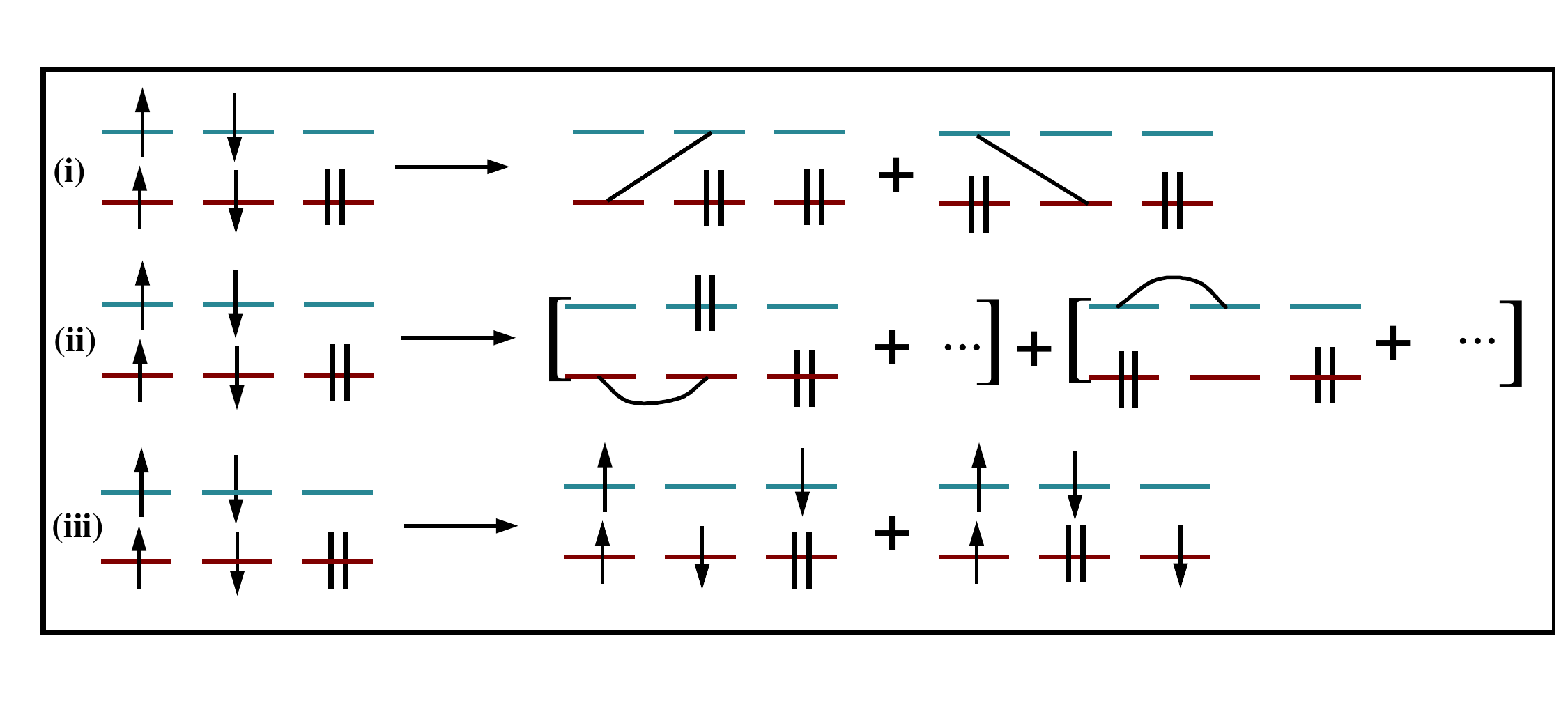}
\caption{\label{fig:ESA-processes} Schematic representation of the intermonomer CT channels responsible for the labeled ESA peaks in Fig.~\ref{fig:ESA-spectra}. Processes (i) and (ii) occur in both 2Pc and longer oligomers. Process (ii) corresponds to the TA band monitored in Ref.~\onlinecite{Sanders16d}. Process (iii) requires additional pentacene units and is therefore allowed only for $n>2$.}
\end{figure}

TA is due to ESAs from the initial excited state. ESAs that originate from $^1$(T$_1$T$_1$) and are absent in T$_1$ monomer spectra
can only be due to intermonomer CT between neighboring singlet spin-coupled T$_1$ monomers, as 
shown schematically in Fig.~\ref{fig:ESA-processes}. These include, in order (see Fig.~\ref{fig:ESA-processes}), (i) CT from singly-occupied LUMO (HOMO) 
of monomer 1 to HOMO (LUMO) of monomer 2, resulting in a final state that is in the same energy 
manifold as 1e-1h states that contribute to absorption band S$_2$ in Fig.~3, (ii) CT from singly-occupied LUMO (HOMO) of monomer 1 to LUMO (HOMO) of monomer 2, resulting in a high energy
2e-2h eigenstate, and (iii) CT between a T$_1$ monomer and a neighboring monomer in its ground state. 
ESA corresponding to CT process (iii) will not occur in 2Pc, while ESA
due to CT process (i) was a theoretical prediction at the time the experiments of Ref.~\onlinecite{Sanders16d} were performed (the prediction would be confirmed experimentally
later, see Refs. \onlinecite{Trinh17a,Pun19a}). ESA corresponding to process (ii) therefore corresponds to the experimental TA at 710 nm, and its continued persistence in $n$Pc with $n >2$ and PPc
\cite{Sanders16d} therefore led to the idea that the triplet binding energy of the $^1$(T$_1$T$_1$) states in $n$Pc and PPc is forbiddingly high. This interpretation is based on conceptualization Fig.~1(i) of triplet-triplet states. The $^1$(T$_1$T$_1$) eigenstates obtained by us in Figs. 5 and 6 from the PPP calculations
do not require this interpretation, as we show below.

In Fig.~8 we have shown our calculated ESA spectrum for 2Pc, and the individual ESAs from the different $^1$(T$_1$T$_1$) eigenstates of 3Pc (see Fig.~5). The absorption bands 
were as usual calculated using Lorentzians and uniform linewidths of 0.1 eV (see SM V). We assign the 
experimental TA features at ~710 nm to peaks (ii) in the calculated spectra. The computed ESA energies in 3Pc ranges from 1.92 to 1.97 eV (Fig.~5) compare favorably with the experimental TA energy of $\sim$ 1.75 eV. 
The MRSDCI computations of the ESA spectra from the
4Pc eigenstates with its very large basis space is prohobitively expensive timewise, and we have therefore shown the ESA absorption only from a single eigenstate, 
$^1$(T$_1$T$_1$)$_{1[4Pc]}$ of Fig.~6. In this case the computed ESA energies are 0.74 eV, 1.71 eV, 
and 1.97 eV, the highest of which corresponds to peak (ii).
From the wavefunctions of the $^1$(T$_1$T$_1$)$_{2[4Pc]}$ and $^1$(T$_1$T$_1$)$_{3[4Pc]}$, we
expect the absorption spectra of the latter two to be very similar to that of $^1$(T$_1$T$_1$)$_{1[4Pc]}$.

\begin{figure}[h]
\centering{}
\includegraphics[scale=0.32]{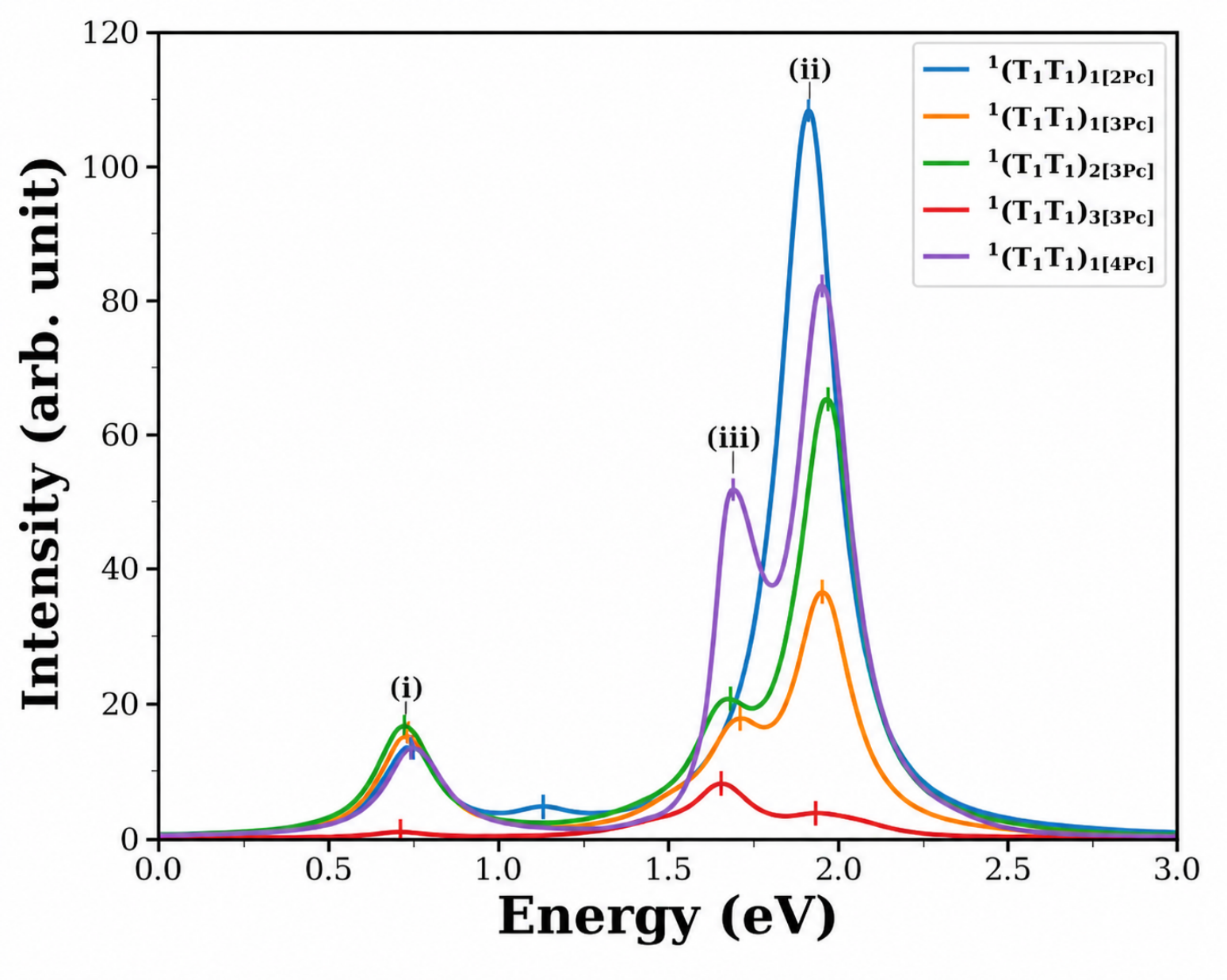}
\caption{\label{fig:ESA-spectra} Calculated triplet-triplet ESA spectra from 2Pc and from selected $^1$(T$_1$T$_1$) eigenstates of $n$Pc. The labels on the peaks correspond to the intermonomer 
CT processes shown in Fig.~\ref{fig:ESA-processes}.}
\end{figure}

The spectra in Fig.~\ref{fig:ESA-spectra} show that the high-energy ESA (ii) seen in 2Pc remains strong for $^1$(T$_1$T$_1$) eigenstates of 3Pc and 4Pc that contain 
substantial nearest-neighbor triplet-pair amplitude. The near degeneracy of the $^1$(T$_1$T$_1$) eigenstates in long oligomers suggest that any of these could be reached
from the optical singlet and they contribute to the TA. This result provides the key interpretation of the experimental TA data. The persistence of a long-lived ESA at the energy of process (ii) in $n>2$ oligomers does not require a picture in which the triplets are trapped in a single localized nearest-neighbor configuration. Instead, it follows naturally from the quantum-superposition character of the $^1$(T$_1$T$_1$) eigenstates shown in Figs.~\ref{fig:TT-P3} and \ref{fig:TT-P4}: even when a state contains contributions from several intertriplet separations, the nearest-neighbor components continue to carry strong oscillator strength for ESA (ii). 


\subsection{The Polymer Limit.}

Understanding the polymer limit in the present case is simple from the exciton basis perspective. As seen from Figs.~4-6, intermonomer
CT in all relevant wavefunctions is limited between nearest neighbor monomers, - next nearest neighbor CT is vanishingly small. 
This is expected,
as the intermonomer $\pi$-electron hopping is only through the terminal C-atoms of the monomers, whose contributions to the 
localized exciton basis MOs is crudely proportional to $1/N^{1/2}$, where N = 26 is the number of C-atoms per monomer. This makes intermonomer hopping very small relative to all other parameters of the oligomer PPP Hamiltonian
and in turn implies that even as the number of eigenstates of each kind increases with oligomer length, their
energies and wavefunction characteristics have already converged for all practical purposes at $n=4$. With increasing $n$ there is
increase in oscillator strength of the ground state optical transitions and concentration of the same to the
translational momentum $k=0$ state of the S$_1$ optical exciton, but no other significant change. This is confirmed from 
examining the experimental $n$-dependent optical absorption spectra that includes the spectrum from the polymer (Fig.~4 in SI of
Ref.~31), - apart from
small red shifts there is little to no difference between the spectra of 3Pc, 4Pc and PPc.
The $^1$(T$_1$T$_1$) eigenstates are expected to be even more localized because of their covalent characters. No significant difference
in the TAs from these states are expected and found (see Fig.~3 in SI of Ref.~31).

\section{Conclusion}

In conclusion, spin-singlet triplet-triplet eigenstates of $n$Pc and PPc are quantum superpositions of $^1$(T$_1$T$_1$) configurations with all possible intertriplet
separations. The complete set of $^n$C$_2$ triplet-triplet eigenstates form a quasi-continuum of nearly degenerate states that occur either below the optical S$_1$ exciton states or are nearly degenerate with these.
The increase in SF rate with $n$ \cite{Sanders16d} is to be anticipated, simply because the number of nearest neighbor charge-charge as well as spin-spin 
interactions increases with size. The experimentally measured similar lifetimes of the TA due to nearest-neighbor $^1$(T$_1$T$_1$) configurations \cite{Sanders16d} is 
also to be anticipated from the electronic structures of the triplet-triplet eigenstates we have determined for $n$Pc (Figs.~5 and 6). 
This is not an indicator of large triplet-triplet
binding energy E$_b$ per se. Nevertheless, triplet separation leading to nearly free triplets occupying entirely distant Pc monomers, as was the original
experimental target \cite{Sanders16d}, are precluded by the electronic structures of the triplet-triplet eigenstates.

The overall excited state electronic structure of PPc and long polyenes have some similarities but the differences are stronger in that CT plays a stronger role in the electronic structures of the triplet-triplet eigenstates the latter. In particular, higher energy nominally triplet-triplets can occur even above the
1$^1$B$_u$ in polyenes. This reraises the question of the feasibility of SF in PPc. While our work precludes straightforward triplet separation along the oligomer
or polymer chain, an
alternate route to SF could be via triplet-triplet states in which the individual triplets occupy neighboring polymer chains, viz., intermolecular as
opposed to intramolecular SF. This is suggested by the
observation of free triplets following photoexcitation in PTc film but not in solution \cite{Pun17a}. This last result is reminiscent of experimental work that has shown that SF in carotenoids is likely not intramolecular \cite{Wang11c,Musser15a,Sutherland23a}. 
The latter result was suggested
theoretically in Ref.~\onlinecite{Aryanpour15a}, although from considerations different from that in the experimental papers. Differences between the triplet-triplet states in polyenes and
PPc and PTc notwithstanding, there is one similarity, viz., in neither case there occurs an energetically distinct triplet-triplet eigenstate consisting  
entirely of many-electron configurations with distant triplet-triplets. 
This last condition appears to be a requirement for successful 
triplet separation, as was
demonstrated experimentally in later experimental work on pentacene-(tetracene)$_n$-pentacene oligomers, in which near total triplet separation occurs with the triplet
excitations occupying the terminal pentacene monomers \cite{Pun19a}. Theoretically, it has been shown that the single many-electron configuration 
with triplets occupying the terminal pentacenes is indeed a unique quantum mechanical eigenstate in these latter compounds \cite{Chesler24a,Nazir25a}.

\section*{Supplementary Material}
See the supplementary material for the MRSDCI configuration-space dimensions, representative excited-state absorption wave functions, and oscillator strengths used in the analysis.

\section*{Acknowledgments}
Work at the University of Arizona was partially supported by NSF Grant No. NSF-DMR-2301372. Some of the calculations were performed using High Performance
Computing Resources maintained by the University of Arizona Research Technologies department and supported by the University of Arizona Technology and
Research Initiative Fund, University Information Technology Services, and Research, Innovation and Impact.

Rupali Jindal: Formal analysis (lead); Investigation (lead); Visualization (lead); Writing - original draft (supporting); Writing - review and editing (supporting). Alok Shukla: Methodology (supporting); Supervision (supporting); Writing - review and editing (supporting). Sumit Mazumdar: Conceptualization (lead); Formal analysis (supporting); Funding acquisition (lead); Methodology (lead); Supervision (lead); Writing - original draft (lead); Writing - review and editing (lead).

\section*{Data Availability}
The data that support the findings of this study are available from the corresponding author upon reasonable request.

\bibliographystyle{aipnum4-2}
\bibliography{Fission-JCP-Revised}

\end{document}